\documentclass[12pt]{article}

\usepackage{latexsym} 
\usepackage{amssymb}  
\usepackage{amsbsy}   
\usepackage{graphicx}       

\newcommand{\al}{\alpha}

\newcommand{\De}{\Delta}
\newcommand{\ep}{\varepsilon}

\newcommand{\Om}{\Omega}

\newcommand{\p}{\partial}
 
\newcommand{\txt}{\textstyle}

\newcommand{\dsp}{\displaystyle}

\newcommand\eqn[1]{(\ref{#1})}      
\newcommand\Eqn[1]{Eq.~(\ref{#1})}  

\newcommand{\beq}{\begin{equation}}
\newcommand{\eeq}{\end{equation}}
\newcommand{\ba}{\begin{array}}
\newcommand{\bea}{\begin{eqnarray}}
\newcommand{\ea}{\end{array}}
\newcommand{\eea}{\end{eqnarray}}
\newcommand{\bi}{\begin{itemize}}  
\newcommand{\ei}{\end{itemize}}
\newcommand{\ben}{\begin{enumerate}} 
\newcommand{\een}{\end{enumerate}}
\newcommand{\bc}{\begin{center}}
\newcommand{\ec}{\end{center}}
\newcommand{\bl}{\begin{flushleft}}
\newcommand{\el}{\end{flushleft}}
\newcommand{\br}{\begin{flushright}}
\newcommand{\er}{\end{flushright}}

\newcommand\comment[1]{ \hbox{[{\it Comment suppressed here.}\/]} }
\newcommand\hide[1]{}

\renewcommand{\O}{{\cal O}}

\newcommand{\skipover}[1]{}

\newcommand{\half} {{\txt \frac{1}{2}}}
\newcommand{\third}{{\txt \frac{1}{3}}}

\newcommand{\twothirds}{{\txt \frac{2}{3}}}

\makeatletter 

\def\appendix{\par                              
    \setcounter{section}{0}                     
    \setcounter{subsection}{0}
    \renewcommand{\theequation}{\Alph{section}.\arabic{equation}}
    \renewcommand{\thesection}{Appendix \Alph{section}
                \setcounter{equation}{0}  } 
}

\def\applabel#1{\@bsphack
  \protected@write\@auxout{}%
         {\string\newlabel{#1}{{\Alph{section}}{\thepage}}}%
  \@esphack}

\def\section{
\setcounter{equation}{0}        
\@startsection {section}{1}{\z@}{-3.5ex plus -1ex minus 
 -.2ex}{2.3ex plus .2ex}{\large\bf}}
\renewcommand{\theequation}{\arabic{section}.\arabic{equation}}

\def\subsection{\@startsection{subsection}{2}{\z@}{-3.25ex plus -1ex minus 
 -.2ex}{1.5ex plus .2ex}{\normalsize\bf}}

\def\subsubsection{\@startsection{subsubsection}{3}{\z@}{-3.25ex plus
 -1ex minus -.2ex}{1.5ex plus .2ex}{\normalsize}}

\makeatother   

\newsavebox{\eqlabel}

\makeatletter  
\newlength{\numblen}
\newsavebox{\eqnumb}
\def\@eqnnum{\savebox{\eqnumb}{\rm (\theequation)}%
\settowidth{\numblen}{\usebox{\eqnumb}}%
\makebox[\numblen][l]{\usebox{\eqnumb}~~~\usebox{\eqlabel}}}
\makeatother   

\newenvironment{equationwithlabel}[1]{ %
  \begin{equation}\label{#1} }{\end{equation}} 
\newcommand{\beql}[1]{\begin{equationwithlabel}{#1}}
\newcommand{\eeql}{\end{equationwithlabel}}

\newcommand{\fm}{{\rm fm}} 
\newcommand{\km}{{\rm km}} 
 
\newcommand{\MeV}{{\rm MeV}}

\newcommand{\Msolar}{M_\odot} 

\begin{document}

\title{\bf 
Compact stars with color superconducting quark matter
}

\author{
Mark Alford${}^{(a)}$, Sanjay Reddy${}^{(b),(c)}$ 
\\[0.5ex]
\parbox{0.6\textwidth}%
{
\normalsize
\begin{itemize}
\item[${}^{(a)}$] 
  Physics and Astronomy Department, \\
  Glasgow University, \\
  Glasgow~G12~8QQ, U.K.
\item[${}^{(b)}$] 
  Center for Theoretical Physics, \\
  Massachusetts Institute of Technology, \\
  Cambridge, MA~02139, U.S.A.
\item[${}^{(c)}$] 
  Theoretical Division, \\
  Los Alamos National Laboratory, \\
  Los Alamos, NM~87545, U.S.A.
\end{itemize}
}}

\newcommand{\preprintno}{
  \normalsize GUTPA/02/11/01, LA-UR-02-7218
}

\date{Nov 13, 2002 \\[1ex] \preprintno}

\begin{titlepage}
\maketitle
\def\thepage{}          

\begin{abstract}
We study compact stars that contain quark matter.  We look at the
effect of color superconductivity in the quark matter on the
nuclear-quark matter transition density, mass-radius relationship, and
the density discontinuity at the boundary between nuclear and quark
matter.  We find that color superconducting quark matter will occur in
compact stars at values of the bag constant where ordinary quark
matter would not be allowed.  We are able to construct ``hybrid''
stars with a color superconducting quark matter interior and nuclear
matter surface, with masses in the range $1.3$-$1.6~\Msolar$ and radii
$8$-$11~\km$.  Our results are consistent with recent mass-radius
limits based on absorption lines from EXO0748-676.
\end{abstract}

\end{titlepage}

\renewcommand{\thepage}{\arabic{page}}

\section{Introduction}
\label{sec:intro}

If matter is compressed far enough beyond nuclear density then there is a
transition from nuclear matter to quark matter.  It is becoming widely
accepted that quark matter will typically be in a color-superconducting
phase \cite{Barrois,BailinLove,ARW1,RappETC,CFL}, in which the quarks
near the Fermi surface form Cooper pairs which condense, breaking the color
gauge symmetry (for reviews see Ref.~\cite{Reviews}).  
The pairing pattern favored at sufficiently high density is
the color-flavor locked (CFL) phase in which up-down, down-strange, and
up-strange Cooper pairs all form, allowing quarks of all three colors and all
three flavors to pair~\cite{CFL}.

One of the most likely locations for quark matter in nature
is the interior of compact stars, where pressure due to gravity
drives the density above nuclear density, and the temperature is
low compared with nuclear/quark energy scales. Various possible
signatures of color superconductivity in compact stars have been
studied, mostly focusing on transport properties (for a recent review,
see Ref.~\cite{ichep}). Although the effect of unpaired quark matter
on the compact star mass-radius relationship
is an active area of research 
\cite{Burgio:2002sn,Schertler:2000xq,Steiner:2000bi},
the consequences of color superconducting quark matter have
not yet been investigated.

The contribution of color superconductivity with gap $\De$ to the
pressure $p$ of quark matter is of order $\mu^2\De^2$, which is
dominated by the leading order $\mu^4$ contribution from the Fermi
sea.  However, quark matter must also pay a free-energy cost, the bag
constant $B$, relative to the confined vacuum: 
\beq p \sim \frac{3}{4\pi^2} \mu^4 +
\frac{3}{\pi^2} \De^2\mu^2 - B~.  
\eeq 
If the bag constant is large enough so that nuclear matter and quark
matter have comparable pressures at some density that occurs in
compact stars, then the superconducting gap $\De$ may have a large
effect on the equation of state and hence on the mass-radius
relationship of a compact star. A similar observation has been
used recently to show that the region of model space where strange
quark matter is absolutely stable is influenced by
superconductivity \cite{Lugones:2002ak}.

In this paper we study the gross structure of compact stars, taking
into account the possibility of a CFL quark-paired phase of quark
matter, as well as unpaired quark matter (UQM).  We treat the quark
matter as a Fermi sea of free quarks with an additional contribution
to the pressure from the formation of the CFL condensate.  We treat
the nuclear matter as consisting only of protons, neutrons, and
electrons, and either obtain the nuclear equation of state from the
Walecka model of nuclear interactions, or use the APR98 equation of
state \cite{APR98}.

In order to find the effects of color superconductivity, we allow all
the phases to compete with each other, selecting the highest pressure
phase at each value of the quark chemical potential.  If local
electrical neutrality is imposed, this leads to sharp interfaces
between the different phases. If a globally neutral inter-penetration
of two charged phases is allowed, this leads to mixed phases. We study
both possibilities.  We survey the parameter space of the quark
matter, covering a range of values of the bag constant $B$, the
strange quark mass $m_s$, and the color superconducting gap $\De$.  In
section \ref{sec:EoS} we explain our calculation of the equations of
state for nuclear and quark matter.  In section \ref{sec:structure} we
obtain the resultant mass-radius relationships, paying particular
attention to the maximum masses that can be obtained. Section
\ref{sec:conclusions} presents our conclusions.

\section{Equations of state}
\label{sec:EoS}

For nuclear matter we use the Walecka equation of state, which allows
us to calculate the pressure for any quark chemical potential $\mu$
and electron chemical potential $\mu_e$, and hence to construct mixed
phases. In section \ref{sec:maxmass} we will also show results for the
APR98 equation of state, which is obtained using non-relativistic
variational methods starting from a Hamiltonian that reproduces known
nucleon-nucleon scattering data.
For nuclear matter at very low densities we use the tabulated
Negele-Vautherin \cite{NV}
and Baym-Pethick-Sutherland \cite{BPS} equations of state.

\subsection{Walecka equation of state}
\label{sec:Walecka}

We use the Walecka model as described in \cite{GBOOK} and calibrated
in \cite{SW}. The free energy density is
\begin{eqnarray}
\Omega_{\rm nuclear}(\mu_n,\mu_e)&=& \frac{1}{\pi^2}\left(
\int_0^{k_{Fn}}dk~k^2~(\ep_n(k) - \mu_n) + 
\int_0^{k_{Fp}}dk~k^2~(\ep_p(k) - \mu_p)  
\right) \,\nonumber \\
&+&\frac{1}{2}\left(m_{\sigma}^2\sigma^2 
- m_{\omega}^2\omega^2-m_{\rho}^2\rho^2\right)+U(\sigma)
- \frac{\mu_e^4}{12\pi^2} \ , 
\label{omeganuc}
\end{eqnarray}
where
\begin{eqnarray}
\ep_n(k) &=& 
  \sqrt{k^2+{m^\ast_N}^2} +g_{\omega N} \omega 
  - \half g_{\rho N}\rho \,,\\
\ep_p(k) &=& \sqrt{k^2+{m^\ast_N}^2} +g_{\omega N} \omega 
  + \half g_{\rho N}\rho \,,
\end{eqnarray}
are the neutron and proton single particle energies in the mean field
approximation.  The corresponding Fermi momenta
$k_{Fn}$ and $k_{Fp}$, which minimize the 
free energy
at fixed baryon and electron chemical potentials, are given by
solving
\beq
\label{walecka1}
\ba{rcl}
\ep_n(k_{Fn}) &=& \mu_n \ ,\\
\ep_p(k_{Fp}) &=& \mu_p \ ,
\ea
\eeq
where weak interaction equilibrium sets $\mu_p=\mu_n-\mu_e$, and 
\beq
\label{walecka2}
\ba{rcl}
m_\sigma^2  \sigma &=& 
  \dsp  g_{\sigma N} \left(\rule[-0ex]{0em}{2ex}
  n_s(k_{Fn}) + n_s(k_{Fp})\right) 
  - \frac{dU}{d\sigma} \ , \\[1ex]
m_\omega^2  \omega &=& g_{\omega N} \left(\rule[-0ex]{0em}{2ex}
  n(k_{Fn})+n(k_{Fp})\right) \ , \\[1ex]
m_\rho^2  \rho &=&  \half g_\rho \left(\rule[-0ex]{0em}{2ex}
  n(k_{Fp})-n(k_{Fn})\right) \ . 
\ea
\eeq
The nucleon number density $n$ and scalar density $n_s$
for nucleons
with Fermi momentum $k_F$ are
\beq
\ba{rcl}
n(k_F) &=&  \dsp \frac{1}{\pi^2} \int_0^{k_F} dk\, k^2 = 
  \frac{k_F^3}{3\pi^2} \ , \\[2ex]
n_s(k_F) &=& \dsp \frac{1}{\pi^2}
  \int_0^{k_F} dk \, k^2 \frac{m^\ast_N}{\sqrt{k^2 + {m^\ast_N}^2}} \ ,
\ea
\eeq
\hide{
\begin{eqnarray}
k_{Fn}&=&\sqrt{(\mu_n - g_{\omega N} \omega 
  - \half g_{\rho N}\rho)^2 - {m^\ast_N}^2} 
\,
\label{kfn} \\
k_{Fp}&=&\sqrt{(\mu_n - \mu_e - g_{\omega N} \omega + \half g_{\rho N}\rho)^2 
- {m^\ast_N}^2}
\label{kfp}
\,,
\end{eqnarray}
}
where 
\beq
m^\ast_N = m_N-g_{\sigma N}\sigma
\eeq is the nucleon effective mass, which
is reduced compared to the free nucleon mass $m_N$ due to the scalar field 
$\sigma$, taken to have $m_\sigma=600$~MeV.
The scalar self-interaction term is 
\begin{equation}
U(\sigma)= \frac{b}{3}m_N(g_{\sigma N}\sigma)^3 + \frac{c}{4}
(g_{\sigma N}\sigma)^4\ ,
\end{equation}
where $b$ and $c$ are
dimensionless coupling constants.
The five coupling constants,
$g_{\sigma N}$, $g_{\omega N}$, $g_{\rho N}$, $b$, and $c$, are chosen 
as in Ref.~\cite{SW} to reproduce
five empirical properties of nuclear matter at saturation density:
the saturation density itself is $n_0 =0.16{\rm ~fm}^{-3}$;
the binding energy per nucleon is 16 MeV; the nuclear compression
modulus is 240 MeV; the nucleon effective mass at saturation
density is $0.78 m_N$;
and the symmetry energy is 32.5 MeV.

The charge density in nuclear matter is
\beq
Q_{\rm nuclear} = \frac{\p \Om_{\rm nuclear}}{\p \mu_e}
\label{Walecka_charge}
\eeq
which is just the number density of protons minus that of electrons.
In bulk matter one requires $Q_{\rm nuclear}=0$, but not in mixed phases
(Section~\ref{sec:mixed}).

\subsection{Unpaired quark matter (UQM) equation of state}
\label{sec:unpaired}

In noninteracting unpaired quark matter, and neglecting the light
quark masses, the free energy density is 
\beq \ba{rl} \Om_{\rm
UQM}(\mu,\mu_e) = &\dsp \phantom{+}
\frac{3}{\pi^2}\int_0^{\nu_u}p^2(p-\mu_u)dp +
\frac{3}{\pi^2}\int_0^{\nu_d}p^2(p-\mu_d)dp \\ [3ex] &\dsp +
\frac{3}{\pi^2}\int_0^{\nu_s}p^2\left(\sqrt{p^2+m_s^2}-\mu_s\right)dp\,
\ea \eeq where the Fermi momenta are \beq \ba{rcl} \nu_u^2 &=&\mu_u^2-
m_u^2 \quad {\rm where}~\mu_u=\mu-\twothirds \mu_e\ , \\ [1ex] \nu_d^2
&=& \mu_d^2 - m_d^2 \quad {\rm where}~\mu_d=\mu+\third \mu_e\,\\ [1ex]
\nu_s^2 &=& \mu_s^2 - m_s^2 \quad {\rm where}~\mu_s=\mu-\third \mu_e\
.  \ea \eeq Differentiating with respect to $\mu_e$, we obtain the
charge density \beq Q_{\rm UQM} = \frac{2~\mu^2\mu_e}{\pi^2} -
\frac{2~\mu}{3\pi^2}~(\mu_e^2+\frac{3}{4}m_s^2)-\frac{m_s^2
\mu_e}{6\pi^2} +
\O \left[\frac{m_s^4}{\mu},\frac{m_s^4\mu_e}{\mu^2}\right] \ .
\label{UQM_charge}
\eeq

\subsection{CFL quark matter equation of state}
\label{sec:CFL}

We describe the CFL phase using the free energy
\beq
\Om_{\rm CFL}(\mu,\mu_e) = \Om^{\rm quarks}_{\rm CFL}(\mu)
+ \Om^{\rm GB}_{\rm CFL}(\mu,\mu_e) + \Om^{\rm electrons}(\mu_e)
\label{OmegaCFL}
\eeq
The contribution to \eqn{OmegaCFL}
from the quarks is \cite{Reviews,neutrality}
\begin{equation}
\Omega_{\rm CFL}^{\rm quarks} = \frac{6}{\pi^2}\int_0^{\nu}p^2(p-\mu)dp + 
\frac{3}{\pi^2}\int_0^{\nu}p^2\left(\sqrt{p^2+m_s^2}-\mu\right)dp
-\frac{3 \Delta^2\mu^2}{\pi^2} 
+ B\ ,
\label{OmegaQuark}
\end{equation}
where the quark number densities 
are $n_u=n_d=n_s=(\nu^3+2\Delta^2\mu)/\pi^2$
and the common Fermi momentum is 
\begin{equation}
\nu=2\mu-\sqrt{\mu^2+\frac{m_s^2}{3}}
\end{equation} 
The first two terms give the free energy of the noninteracting quarks, while
the third term is the lowest order (in powers of $\Delta/\mu$) contribution
from the formation of the CFL condensate.

The contribution to \eqn{OmegaCFL} from the Goldstone bosons arising
due to breaking of chiral symmetry in the CFL phase is denoted
$\Omega^{\rm GB}_{\rm CFL}(\mu,\mu_e)$. The effective theory
describing the octet of mesons has been studied extensively in earlier
works \cite{effectivetheory}. When the electron chemical potential
exceeds the mass of the lightest negatively charged meson, which in
the CFL phase is the $\pi^-$, these mesons condense
\cite{BedaqueSchaefer,KaplanReddy}. The free energy in this case is
given by
\begin{equation}
\Omega^{\rm GB}_{\rm CFL}(\mu,\mu_e)=-\frac{1}{2}f_{\pi}^2 \mu_e^2
 \left(1-\frac{m_{\pi}^2}{(\mu_e)^2} \right)^2 \ ,
\end{equation}
where the parameters are \cite{effectivetheory}
\beq
f_\pi^2 = \frac{(21 - 8 \ln 2)\mu^2}{36\pi^2}, \qquad
m_{\pi^-}^2 = \frac{3 \De^2}{\pi^2f_\pi^2 } m_s(m_u+m_d)\,.
\label{pion_params}.
\eeq
Finally, the contribution to \eqn{OmegaCFL} from electrons is
\beq 
\Om^{\rm electrons}(\mu_e) = - \frac{\mu_e^4}{12\pi^2}\ .  
\eeq

Unlike the UQM phase, in the CFL phase there is a gap in the quark
excitation spectrum, and the lightest charged excitations correspond
to pions and kaons. The charge susceptibility in this phase is
determined by the effective theory for these collective modes. The
electric charge density (carried by the pion condensate) induced by a
electron chemical particle is given by \cite{KaplanReddy}
\beq Q_{\rm CFL}= -f_{\pi}^2
\mu_e \left[1 - \frac{m_{\pi}^4}{\mu_e^4}\right] \ .
\label{CFL_charge}
\eeq

Meson condensation can occur in the CFL phase even in the absence of
an electric charge chemical potential. Bedaque and Schaefer
\cite{BedaqueSchaefer} have shown that the strange quark mass
introduces a stress on the CFL state and that might result in the
condensation of $K^0$ mesons. Condensation occurs when $m_s^2/2\mu \ge
m_{K^0}$, where $m_{K^0}$ is the mass of the $K^0$ meson in the CFL
phase. The free energy due to $K^0$-condensed phase is
\begin{equation}
\Omega^{\rm GB}_{\rm CFL}(\mu)=-\frac{1}{2}f_\pi^2 \frac{m_s^4}{4 \mu^2}
 \left(1-\frac{4\mu^2 m_{K^0}^2}{m_s^4}\right)^2 \ ,
\label{omegakzero}
\end{equation}
where the kaon mass \cite{effectivetheory}
\beq
m_{K^0}^2 = \frac{3 \De^2}{\pi^2f_{\pi}^2 } m_u(m_d+m_s)\,.
\label{kaon_params}
\eeq 

{}From Eq.~\eqn{omegakzero} we see that the free energy due to
$K^0$-condensation is an order $m_s^4$ effect and thereby small
compared to the $\De^2 \mu^2$ contribution to the free energy for $\De
\sim 100$ MeV. For this reason we neglect $K^0$ condensation in this
study.

\subsection{Is color superconductivity important for bulk structure?}
We see from \eqn{OmegaQuark} that color superconductivity contributes
$\O(\mu^2\De^2)$ to the free energy, which is small relative to the
kinetic energy density which is $\O(\mu^4)$.  This well-known
suppression is a consequence of the fact that pairing is a Fermi
surface phenomena and the superconducting gap is usually small
compared to the chemical potential.  Naively, this would lead us to
conclude to that superconductivity will not greatly affect the
equation of state of quark matter. If this were true, we should expect
that the mass-radius relation of neutron stars containing
superconducting quark matter would be nearly identical to those
constructed in earlier works wherein the role of superconductivity was
neglected. However, the situation is more complicated.  In the bag model
description of quark matter, the free energy gets an additional
contribution due to the bag constant. The kinetic pressure and bag
pressure cancel when the quark chemical potential has value
\begin{equation}
\mu_0= \left(\frac{4\pi^2~B}{3}\right)^{\frac{1}{4}} 
 + \O \Bigl(\frac{m_s^2}{\mu}\Bigr)\,.
\end{equation}
Thus, for a given $B$, there is a narrow window in
quark chemical potential in which the pairing contribution to the pressure
is dominant. In the vicinity of $\mu_0$,  superconductivity will 
therefore make a significant contribution to the equation of state of 
quark matter.  For $B^{1/4}$ in the range $150$-$200$ MeV 
($B = 66$-$210~\MeV\!/\fm^3$), we find $\mu_0 \simeq 320-400$ MeV. 
This is an interesting range of
chemical potentials because the phase transition from nuclear matter to quark
matter typically occurs here. Further, and perhaps more importantly, 
the pairing contribution to the pressure of CFL quark matter, 
$P_{\Delta}=3 \Delta^2 \mu^2 / \pi^2$, can be
comparable or larger than the pressure in the nuclear phase at the same baryon
chemical potential. Superconductivity will thereby significantly
influence the critical chemical potential at which the transition from 
nuclear to quark matter occurs. This is clear from Fig.~\ref{fig:mixed_eos},
where the transition from NM to QM occurs at a much lower pressure for
CFL QM ($a\!\to\! b$) than for unpaired QM ($c\!\to\! d$).
The critical chemical
potential for the transition for different values of the bag constant and 
the superconducting gap are shown in table \ref{tab:transition}.

\begin{table}
\begin{tabular}{rrrrrrr}
\hline
\multicolumn{2}{c}{~~~~Bag constant} & CFL & chemical & transition
  & nuclear & CFL~~ \\
$B^{1/4}$ & $B$~~~~
   & gap & potential & pressure & density & density \\
 (MeV) &  ($\MeV\!/\fm^3$) & (MeV) & (MeV) 
   & ($\MeV\!/\fm^3$) &($n_{\rm sat}$) & ($n_{\rm sat}$) \\
\hline
190  & 169.6  &   0   & 422.5 & 111.5 & 3.464 & 5.833 \\
190  & 169.6  &   50  & 408.2 & 88.68 & 3.158 & 5.405 \\
190  & 169.6  &   100 & 365.4 & 33.79 & 2.149 & 4.295 \\
170  & 108.7  &   0   & 352.6 & 21.62 & 1.805 & 3.297 \\
170  & 108.7  &   50  & 338.4 & 10.77 & 1.382 & 3.033 \\
\hline
\end{tabular}
\caption{
Properties of the nuclear-quark phase transition for various
bag constants $B$ and color-superconducting gaps.
Nuclear matter is treated using the Walecka model.
The size of the gap has a significant effect on the pressure at which
the phase transition occurs and the densities of the two phases there.
}
\label{tab:transition}
\end{table}

\subsection{Mixed phases}
\label{sec:mixed}

In the preceding discussion we have enforced local charge neutrality
in the nuclear and quark phases. We have neglected the possibility of
having a mixed phase, at finite $\mu_e$, containing positively charged
nuclear matter co-existing with negatively charged CFL
matter\cite{Glendenning:1992vb}. Such a possibility was considered in
Ref.~\cite{ARRW} where the bulk free energy difference between the
homogeneous phases and the heterogeneous mixed phase was
calculated. The free energy difference between the homogeneous phase
and the mixed phase was found to be quite small. If one accounted for
the additional surface energy cost in the mixed phase, it was found
that even modest values of the surface tension, $\sigma_{\rm NM-CFL}
\sim 30 $ MeV/fm$^2$ were sufficient to disfavor the mixed
phase. Nonetheless, we consider this possibility in this work for two
reasons. First, the surface tension between the nuclear matter and CFL
matter is poorly known. Second, in the case of unpaired quark matter,
allowing for a mixed phase has been shown to significantly affect the
equation of state over a wide range of pressures and consequently
modify the mass-radius relation. This indicates that the
superconducting case warrants investigation.

The procedure to construct the mixed phase between nuclear matter and
CFL matter was outlined in Ref.~\cite{ARRW}. We follow the same
prescription here but briefly note some salient features which
distinguish the nuclear-CFL mixed phase from the mixed phase between
nuclear matter and unpaired quark matter.  The volume fraction of the
nuclear and quark phases in the mixed phase is determined by the
requirement of global charge neutrality. Denoting the charge density
of the nuclear phase by $Q_{\rm nuclear}$ and the charge density of
the CFL phase as $Q_{\rm CFL}$, the volume fraction of the CFL phase
is
\beq 
\chi = \frac{V_{CFL}}{V} 
     = \frac{Q_{\rm nuclear}}{Q_{\rm nuclear}-Q_{\rm CFL}}\ .
\label{chi} 
\eeq 
The energy density of the mixed phase is the volume-weighted 
average of the individual energy densities and is given by $
\epsilon = \chi \epsilon_{\rm CFL} + (1-\chi) \epsilon_{\rm nuclear}$.
{}From Eq.~\eqn{chi}, we see that, if at fixed electron chemical
potential the negative charge density of the quark phase is large its
volume fraction will be correspondingly smaller. This is important
because the charge susceptibilities of the normal and superconducting
phases are quite distinct. In the normal phase it is easy to furnish
electric charge since there is no gap in the spectrum for quarks. The
electric charge densities of the nuclear matter (Walecka), 
unpaired quark matter, and CFL quark matter phases are given in
Eqs.~\eqn{Walecka_charge}, \eqn{UQM_charge}, \eqn{CFL_charge}.

For typical quark and electron chemical potentials encountered in the
neutron star context one finds that the charge density in the normal
phase is significantly larger. For example, when $\mu=400$ MeV,
$m_s=150$ MeV, $\mu_e=100$ MeV and $m_{K}=30$ MeV, the charge density
in the normal phase $Q_{\rm UQM}=-0.32$ fm$^{-3}$, while in the CFL
phase $Q_{\rm CFL}=-0.09$ fm$^{-3}$ (note that neutral CFL QM has
$\mu_e=0$ \cite{neutrality}, so the charge density in CFL is due to
the $\pi^-$ density induced by $\mu_e$).  From the preceding arguments
this implies that the volume fraction of the CFL phase in the mixed
phase will be significantly larger.  For the same reason, although the
pairing contribution to the free energy itself is small, the equation
of state of the CFL-nuclear mixed phase is considerably softer.

\begin{figure}[htb]
\begin{center}
\includegraphics[width=1.0\textwidth,angle=0]{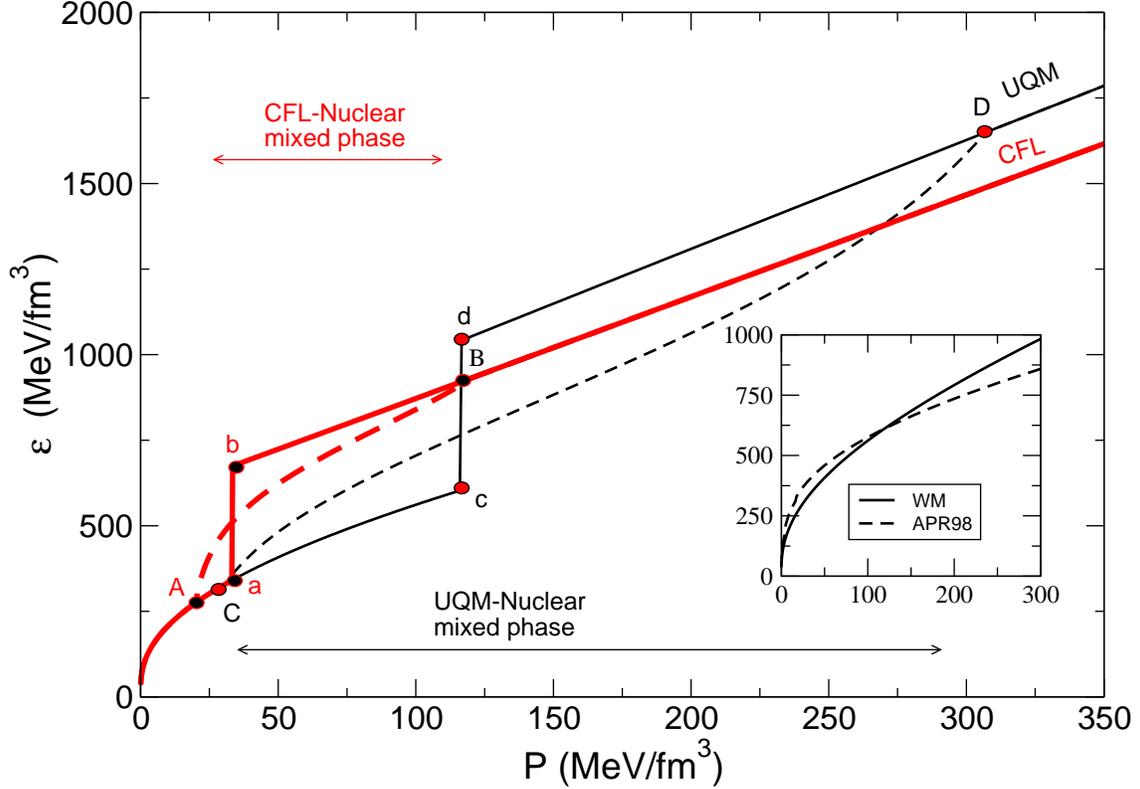}
\end{center}
\caption{ The equation of state for Walecka nuclear matter and
quark matter with $B^{1/4} = 185$ MeV 
($B=153~\MeV\!/\fm^3$)
and $m_s=200$ MeV.  We show
normal unpaired quark matter (thin lines) and color
flavor locked superconducting quark matter with
$\De=100~\MeV$ (thick lines). We show
neutral phases with sharp transitions (solid curves) and mixed phases
(dashed curves).
Differences between the Walecka model equation of state and the APR98 equation of state can be 
inferred from the figure inset.}
\label{fig:mixed_eos}
\end{figure}

The equation of state for the CFL-nuclear and UQM-nuclear mixed phases 
are shown in Fig.~\ref{fig:mixed_eos}. 
We see that if sharp transitions occur, then 
the occurrence of color superconductivity leads to higher energy density at
low pressure because the NM$\to$CFL transition 
($a\!\to\! b$) occurs at lower pressure
than NM$\to$unpaired ($c\!\to\! d$),
but lower energy density at high pressure, because CFL has a 
lower energy density than unpaired QM.
On the other hand, if mixed phases occur, then color superconductivity
leads to a higher energy density up to high pressures, since
the NM+CFL mixed phase ($A\!\to\! B$) and CFL phase that follows it both have
higher energy density
than the NM+unpaired mixed phase ($C\!\to\! D$) over a wide range of pressures.

A mixed phase involving CFL and UQM is also allowed in principle. Such
a mixed phase is characterized by a negatively-charged CFL phase
coexisting with a positively-charged UQM phase. We find that such a
mixed phase has lower free energy (even in the absence of surface and
Coulomb effects) only if the negatively charged mesons (kaons)
condense in the CFL phase. This requires that $\mu_e \ge m_K$. On
the other hand, the UQM phase is positively charged only when $\mu_e
\le m_s^2/4\mu$. Further, the difference in free energy between this
phase and the pure CFL phase is only of order $m_s^4$ and it does not
greatly affect the equation of state. For these reasons we do not consider such a
mixed phase in our study of the structure of the compact
star. Nonetheless, the existence of such a mixed phase might have
important consequences for transport and cooling phenomena especially
in pure quark stars.

\section{Compact Star Structure}
\label{sec:structure}

To determine the mass and radius
of the compact object for a given value of the central pressure,
we must solve the Tolman-Oppenheimer-Volkov (TOV) equations \cite{TOV},
\begin{eqnarray}
\frac{dP}{dr}&=&\frac{-G~M(r)~\epsilon(P)}{r^2~c^2}
\left(1+\frac{P}{\epsilon}\right)
~\left(1+\frac{4\pi r^3 P}{M(r)c^2}\right)
\left(1-\frac{2G M(r)}{r c^2}\right)^{-1} \nonumber \\
\frac{dM(r)}{dr}&=&4\pi^2~\epsilon(P)
\end{eqnarray}
where $P=P(r)$ and the equation of state specifies $\epsilon(P)$,
i.e., the energy density as a function of the pressure, and $M(r)$ is
the total energy enclosed within radius $r$. For a given central
pressure, $P(r=0)$, the above equations can be easily integrated out
to the surface of the star, where $P=0$, to obtain the mass and radius
of the object. By varying the central pressure it is possible to
obtain the mass radius relation predicted for a given model
description of the matter equation of state. The focus of this section
is to employ the equations of state described in the previous section
and deduce the corresponding mass-radius relationship.

\begin{figure}[htb]
\begin{center}
\includegraphics[width=0.6\textwidth,angle=-90]{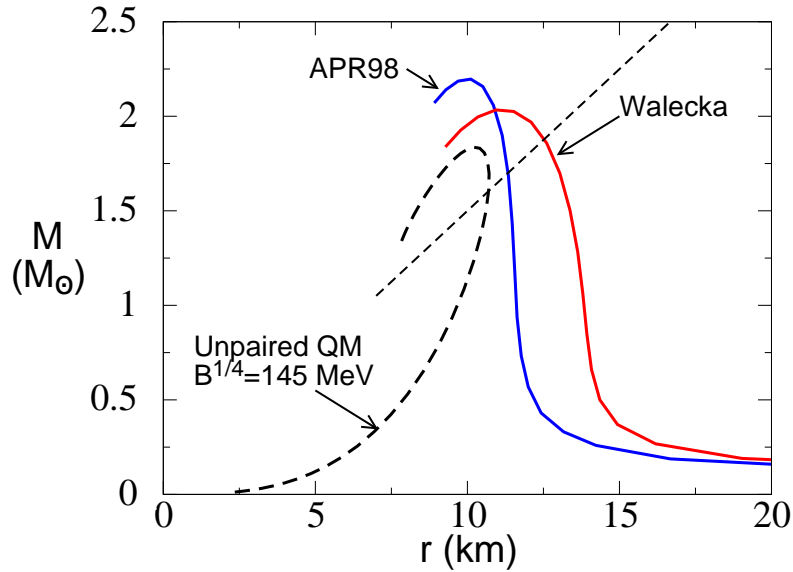}
\end{center}
\caption{ Mass-radius relationships for pure nuclear matter (NM) with
Walecka and APR98 equations of state, and for unpaired quark matter
with $B^{1/4}=145~\MeV$. The straight dashed line indicates the
constraint obtained by recent measurements of the redshift of spectral
lines from EXO0748-676 \cite{Cottam}. 
This constraint requires model equations of state to 
intersect the line $M/R \approx 0.15~(\Msolar/\km)$ }
\label{fig:simpleMR}
\end{figure}

\subsection{$M$-$R$ relationship for uniform stars}

First, we review the mass-radius relationship for the simple cases of
stars made of pure nuclear matter or pure quark matter.  The two solid
curves in Fig.~\ref{fig:simpleMR} are for pure nuclear matter
stars. They follow from solving the TOV equation at high bag constant,
so quark matter is highly disfavored and does not occur.  There is one
for each of the two nuclear equations of state that we use in this
paper: the Walecka model (Section~\ref{sec:Walecka}) and the APR98
tabulation \cite{APR98}.  They are roughly similar, showing the
characteristic sharp rise in mass at $R\approx 10$-$15~\km$ as the NM
equation of state hardens around nuclear saturation density.  They
also show similar maximum masses of about $2~\Msolar$, at central
densities of 5 to 10 times nuclear saturation density. Despite these
similarities there are, as can be seen in the inset of
Fig.~\ref{fig:mixed_eos}, quantitative differences.  The APR98
equation of state is on average softer at low density and stiffer at
high density compared to the mean field model. For a detailed
discussion of how different nuclear equations of state affect 
the mass-radius
relationship of neutron stars see Ref.~\cite{Lattimer:2000nx}.

For comparison, Fig.~\ref{fig:simpleMR} also shows (dashed line) the
$M$-$R$ curve for quark matter with no pairing ($\De=0$) at a low bag
constant ($B^{1/4}=145~\MeV$, $B=58~\MeV\!/\fm^3$) where three-flavor
quark matter is favored over nuclear matter all the way down to zero
pressure, but two-flavor quark matter is less favorable than nuclear
matter at low pressure. The maximum mass is very sensitive to the bag
constant, so the fact that this curve also shows a maximum at around
$2~\Msolar$ is coincidental. (The mass and radius scale as
$1/\sqrt{B}$ \cite{GBOOK}.) Not all the values of $M$ and $R$ that lie
on the curves are stable. The family of stable configurations is
generated by increasing the central pressure, and obtaining an
increasing mass. As soon as the maximum of $M(p_{\rm central})$ is
attained, further increases in $p_{\rm central}$, apparently yielding
lighter stars, will in fact move onto the unstable branch. This means
that the the parts of the $M$-$R$ curves to the left of the maxima in
Figs.~\ref{fig:simpleMR}, \ref{fig:massradius} correspond to stars
that are hydrodynamically unstable to collapse to a black hole.

\subsection{$M$-$R$ relationship for hybrid and color-superconducting stars}

\begin{figure}[htb]
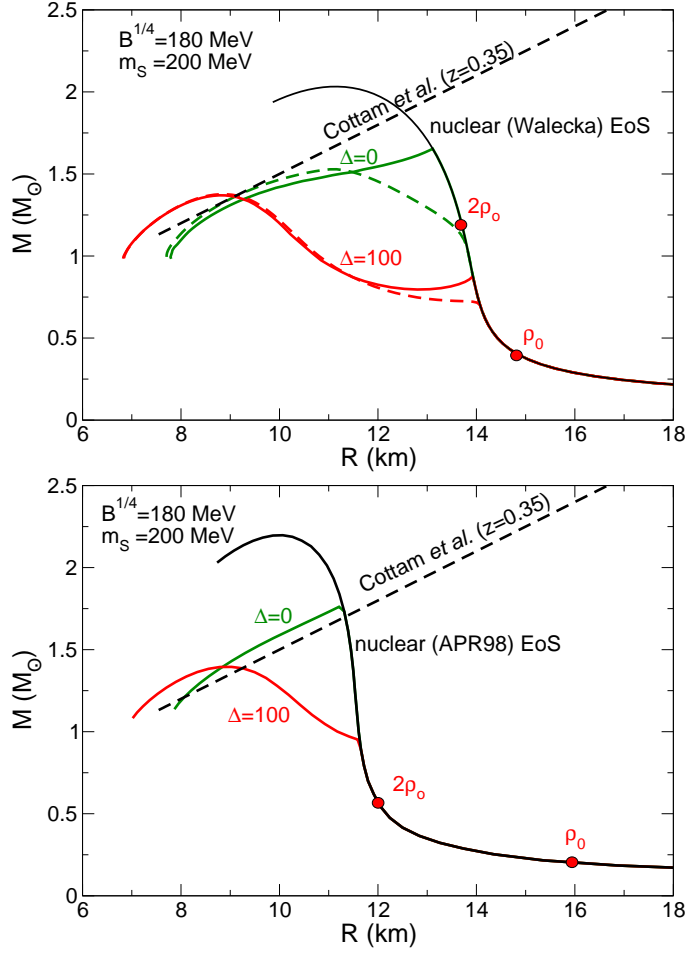

\begin{center}
\includegraphics[width=0.6\textwidth]{figs/mass-radius-wal.eps}
\includegraphics[width=0.6\textwidth]{figs/mass-radius-apr.eps}
\end{center}
\caption{Mass-radius relationships at fixed  bag constant
$B^{1/4}=180~\MeV$ and $m_s=200$ MeV, for unpaired ($\De=0$) 
and color-superconducting ($\De=100~\MeV$) quark matter. 
The mixed phase (dashed) and the sharp
interface curves are shown. The line labeled ``Cottam {\it et al.}''
indicates the constraint obtained by recent measurements of the
redshift on three spectral lines from EXO0748-676 \cite{Cottam}. 
The dots labeled
$\rho_0$ and $2\rho_0$ on the nuclear matter mass-radius curve
indicate that the central density at these locations correspond to
nuclear and twice nuclear saturation density respectively. The 
top panel uses the Walecka equation of state for nuclear matter,
and the lower panel uses APR98 (in which case
 we only consider the sharp-interface scenario).}
\label{fig:massradius}
\end{figure}

The main purpose of this paper is to explore the effect of quark
pairing on the $M$-$R$ relationship at values of the bag constant that
are consistent with nuclear phenomenology.  Fig.~\ref{fig:massradius}
shows the mass-radius curve for a plausible model of dense matter: the
Walecka nuclear equation of state, 
and quark matter with physically reasonable values of
the bag constant $B^{1/4}=180$ MeV 
($B=137~\MeV\!/\fm^3$) and
strange quark mass $m_s=200$ MeV \cite{GBOOK}. 
Curves for unpaired ($\De=0$) and color-superconducting ($\De=100~\MeV$)
quark matter are shown. At these  values
the stars are typically ``hybrid'',
containing both quark matter and nuclear matter.  The solid lines in
Fig.~\ref{fig:massradius} correspond to stars that either have no QM
at all, or a sharp transition between NM and QM: the core is made of
quark matter, which is the favored phase at high pressure, and at some
radius there is a transition to nuclear matter, which is favored at
low pressure.  The transition pressure is sensitive to $\De$, for
reasons discussed earlier. The dashed lines are for stars that contain
a mixed NM-QM phase.  In all cases we see that light, large stars
consist entirely of nuclear matter. When the star becomes heavy
enough, the central pressure rises to a level where QM, either in
a mixed phase or in its pure form, occurs in the core. As can be seen
from the figure the transition density is very sensitive to $\De$.

\begin{figure}[htb]
\begin{center}
\includegraphics[width=0.8\textwidth]{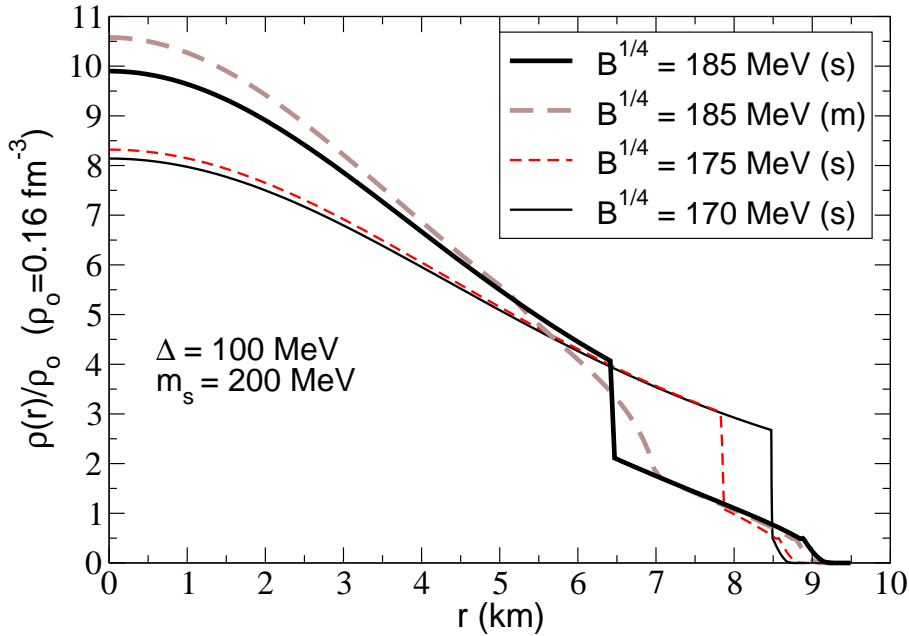}
\end{center}
\caption{Profile of the maximum mass star for bag constant
$B^{1/4}=185,~175,~170~\MeV$ with $m_s=200$ MeV and $\De=100~\MeV$.
The mixed phase (dashed) and the sharp interface curves are shown. The
Walecka model was used to describe the nuclear part of the equation of state.}
\label{fig:profile}
\end{figure}

\clearpage
The profiles of the maximum mass superconducting stars for different
values of the bag constant, $\De=100$ MeV and $m_s=200$ MeV are shown
in Fig.~\ref{fig:profile}. For $B^{1/4}=185$ MeV results for the sharp
interface (denoted as (s)) and the mixed phase (denoted as (m))
scenario are shown. Here the maximum masses correspond to
$1.33~M_{\odot}$ and $1.35~M_{\odot}$, respectively. The maximum mass
for $B^{1/4}=175$ MeV and $B^{1/4}=170$ MeV are
$M_{\max}=1.44~M_{\odot}$ and $M_{\max}=1.52~M_{\odot}$, respectively.
Fig.~\ref{fig:profile} shows that the typical density discontinuity in
the sharp interface scenario is $ \approx 3 \rho_o$. It also shows
that for smaller values of $B$, the 
$NM\!\rightleftharpoons\! QM$
phase transition occurs very
close to the surface of the star (at lower density as discussed
earlier). The denser exterior regions of these stars (despite a less
dense inner core) are primarily responsible for the increase in the
maximum mass observed as one decreases $B$.

Fig.~\ref{fig:profile} indicates that in the mixed phase scenario
there are no discontinuities in the density profile of the
star. However, this is not true in general. It is interesting to note
that even when mixed phases are allowed, there can still be
discontinuities in energy density within them.  In a small range of
parameters, we find stars that have a crust of nuclear matter
surrounding a mixed NM-QM core, but the mixed phase has an outer part
which is a mixture of unpaired QM with NM, and an inner part that is a
mixture of CFL QM with NM. At the interface between the two there is a
density discontinuity {\em within} the mixed phase\footnote{The
inclusion of additional phases, such as the two flavor superconducting
(2SC) phase, might modify this conclusion.}.

The results shown in Figs.~\ref{fig:massradius},\ref{fig:profile}
indicate several generic trends:

\ben \setlength{\itemsep}{-0.5\parskip}
\item The stability of
stars containing pure QM but without a QM-NM mixed phase
(i.e., a sharp-interface scenario) depends on the transition density. Color
superconductivity has a favorable effect on the stability of these
objects since it lowers the transition density and stiffens the
equation of state relative to UQM. 
\item
In general, mixed phase stars are more likely to be stable since the
equation of state (or adiabatic index) changes smoothly in this
case. For example, at $\De=0$ the uniform phase stars become unstable
as soon as quark matter is introduced (the $\De=0$ line slopes down to
the left, showing a decreasing mass as the central pressure rises)
whereas the mixed phase (dashed line) gives a stable branch leading to
a maximum mass $M_{\max}\approx 1.5\,\Msolar$.
\item
For large $\De$ there is very little difference between the sharp
interface and mixed phase mass-radius curves. This is because the
volume fraction of the CFL grows very rapidly within the nuclear-CFl
mixed phase as discussed earlier in section
\ref{sec:mixed}. Consequently, the extent of the mixed phase is
reduced and the equation of state with in this region more closely
resembles the pure CFL equation of state as is evident from
Fig.~\ref{fig:mixed_eos}.
\item 
Although there are visible differences between the mass-radius
curves of the stars with UQM or CFL matter depending on whether one
employs the Walecka or the APR98 equation of state, the maximum mass
and corresponding radius of such a star is fairly independent of the
nuclear equation of state.  This is because the nuclear phase
contributes very little to the total mass of these stars. These
features of the mass-radius relation will be more comprehensively
studied in section \ref{sec:maxmass}.  
\item 
If the transition to superconducting quark matter is constrained to
occur at or above nuclear density the maximum mass of these stars is
$\approx 1.4~M_{\odot}$. It is possible to obtain larger masses
($\approx 1.6~M_{\odot}$) if the transition is allowed to occur at
lower density. We elaborate further on this in the subsequent section.
\een

\subsection{Color superconductivity and the maximum mass}
\label{sec:maxmass}

\begin{figure}[htb]
\begin{center}
\includegraphics[width=0.37\textwidth,angle=-90]{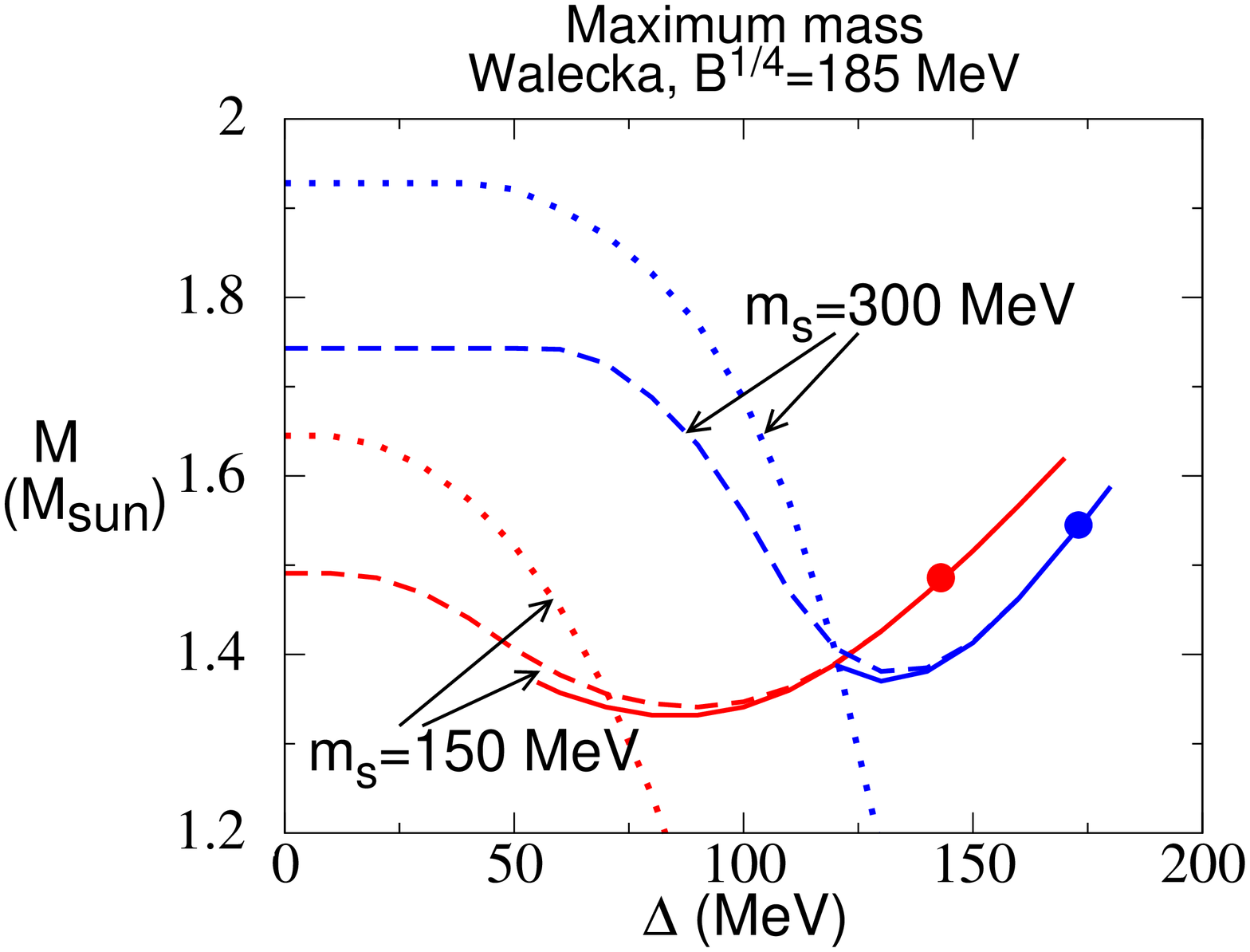}
\includegraphics[width=0.37\textwidth,angle=-90]{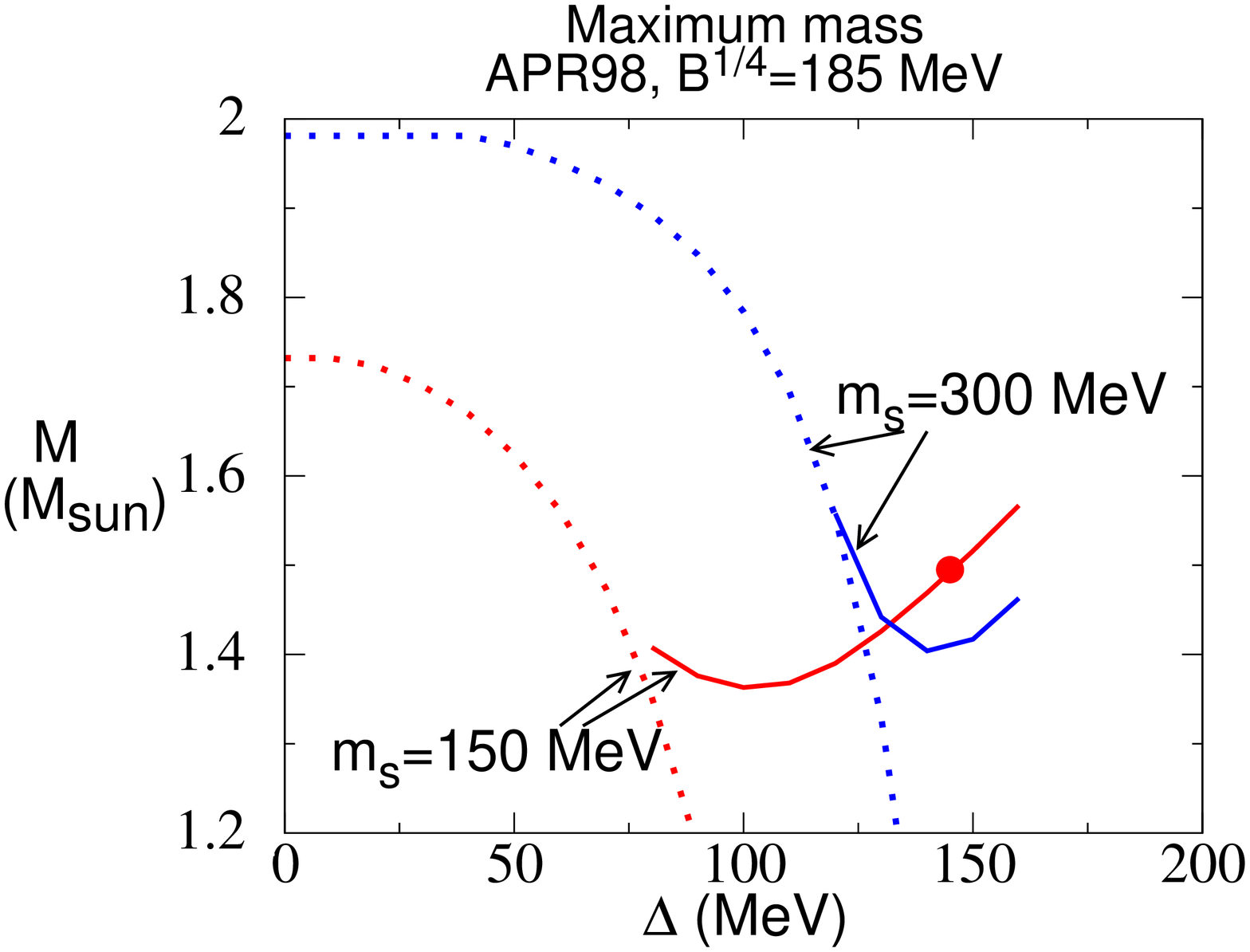}
\includegraphics[width=0.37\textwidth,angle=-90]{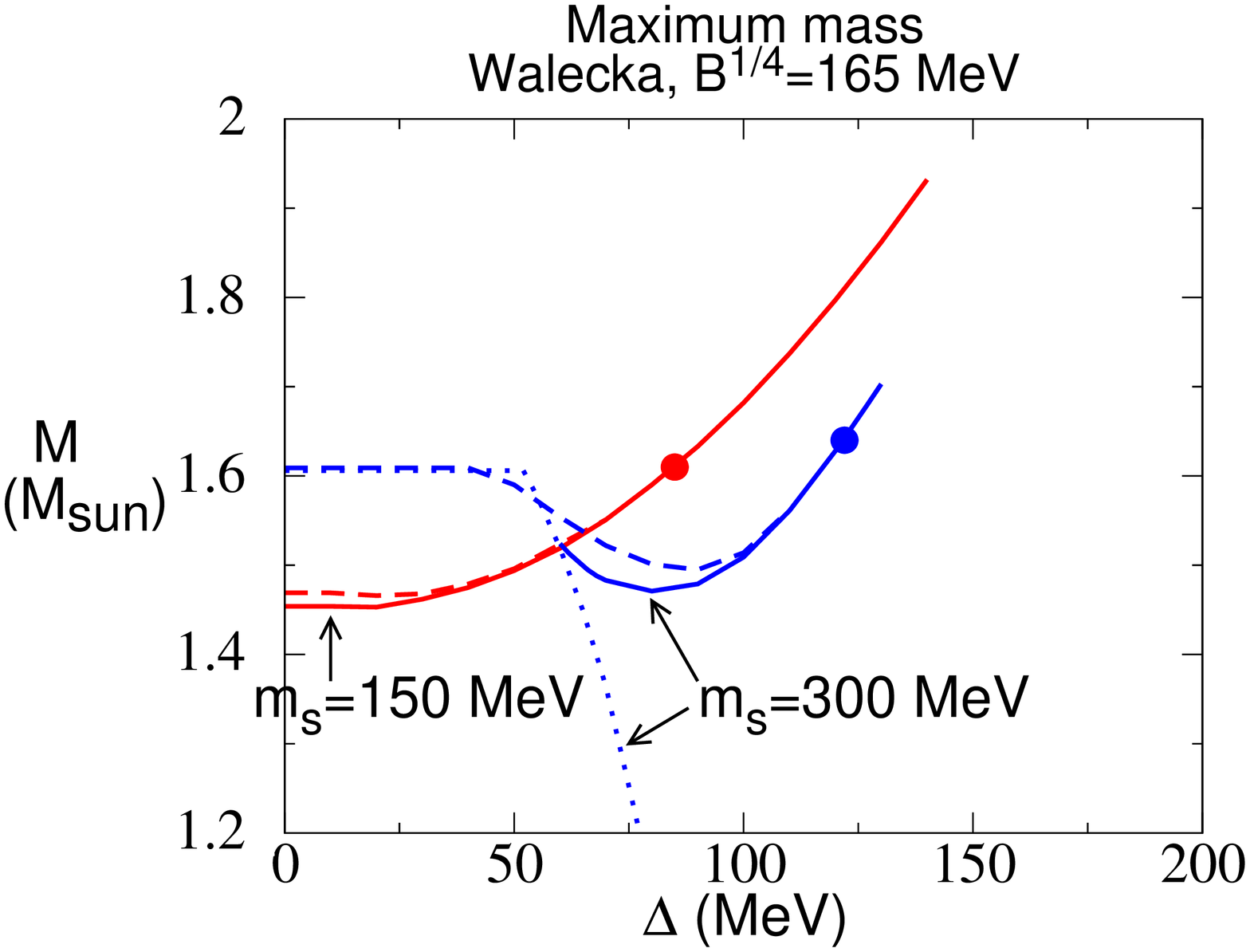}
\includegraphics[width=0.37\textwidth,angle=-90]{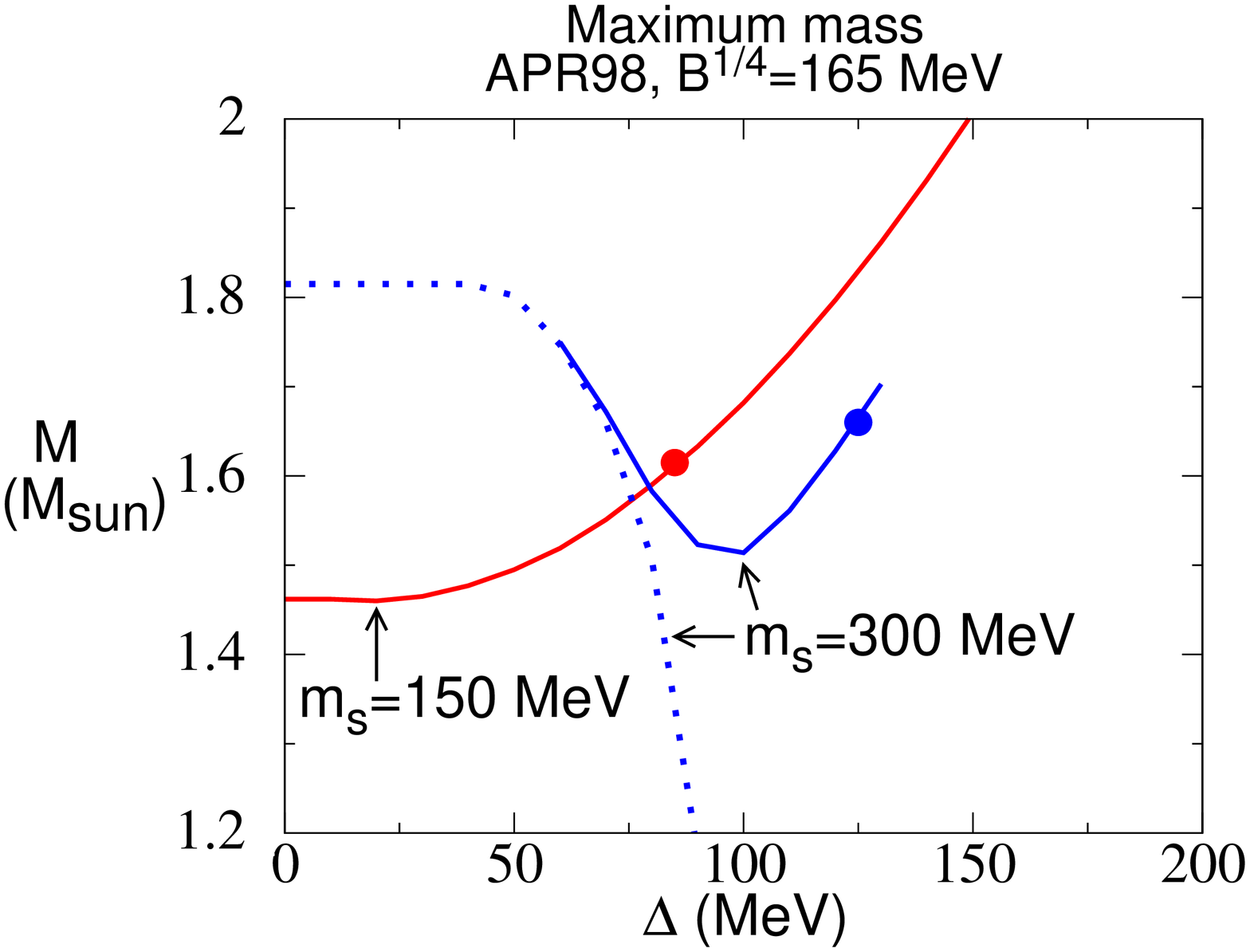}
\end{center}
\caption{ The maximum star mass (in solar masses) attained as a
function of color superconducting gap $\De$.  The upper plots are for
$B^{1/4}=185~\MeV$, the lower plots for $B^{1/4}=165~\MeV$. The left
plots are for nuclear matter described by the Walecka model (in which
case a mixed phase can be constructed).  The right plots are for the
APR98 nuclear equation of state for which we have only constructed
locally neutral phases.  Curves for strange quark mass $m_s=150$ and
$300~\MeV$ are shown.  The solid lines give the radius of stars
with QM only (to the right of the dot), 
or a QM core surrounded by NM (to the left of the dot). 
The dotted lines
show the heaviest pure NM star that occurs at the given gap $\De$.
The dashed lines are for stars that include a mixed phase. 
}
\label{fig:MaxM}
\end{figure}

In bag model treatments of quark matter, the bag constant, strange quark
mass, and the color superconducting gap are unknown parameters.
In this paper we take a reasonable range of values for $B$ and $m_s$,
and study the dependence on $\De$ of observable features of 
compact stars such as their mass and radius.
The resultant predictions can then be used to constrain the
CFL color superconducting gap.  In Fig.~\ref{fig:MaxM} we show how the
maximum star mass (obtained by varying the central pressure) depends
on color superconducting gap $\De$ for
two different bag constants and two different strange quark masses.

For each value of the bag constant and strange quark mass
there are two or three curves of $M_{\rm max}$ vs $\De$.
The solid curve is for stars with some quark matter,
paired or unpaired in them. The prominent dot on the curve
separates the pure quark stars (to the right,
at higher gap) from the stars with
a QM core and a NM crust or mantle (to the left, at lower gap).
The dotted curve indicates the heaviest pure NM star.
This depends on the gap because $\De$ affects the point
in the $M$-$R$ plot at which quark matter appears.
For example, for the equations of state used in the top panel of
Fig.~\ref{fig:massradius}, the maximum NM mass
would be $0.8~\Msolar$, (radius $13.8~\km$), since that is where
QM first appears, and the star ceases to be pure NM.
For the Walecka model that we used, the maximum possible $NM$ mass
is about $2.04~\Msolar$.
For the APR98 equation of state, the maximum possible $NM$ mass
is about $2.20~\Msolar$.
In general, a large gap favors quark matter, causing the heaviest
pure NM star to become lighter.

\vspace{1ex}
\noindent (1) {\bf Large bag constant}, 
$B^{1/4}=185~\MeV$, $B=153~\MeV\!/\fm^3$ \\
(Upper two panels in Fig.~\ref{fig:MaxM}).\\
(a)~Pure NM stars still occur, but their maximum mass drops
as the gap grows (dotted lines in Fig.~\ref{fig:MaxM}) 
as the unstable QM branch cuts off the NM branch.\\ 
(b)~Stars with a QM core and NM surface separated by
a sharp interface (solid lines to the
left of dots) become stable at gap $\De \sim 50$-$120~\MeV$, depending
on $m_s$. These stars have mass $M\lesssim 1.5~\Msolar$.\\
(c)~Stars with a NM-QM mixed phase core and NM surface
(dashed lines) occur at lower values of the gap,
and masses up to around $1.7~\Msolar$ are possible if the strange quark
is heavy enough.\\
(d)~Pure QM stars (solid lines to the right of dots) occur at very large gaps.

\vspace{1ex}
\noindent (2) {\bf Small bag constant}, $B^{1/4}=165~\MeV$,
$B=96~\MeV\!/\fm^3$ \\ (Lower two panels in Fig.~\ref{fig:MaxM}).\\
(a)~Pure NM stars are again cut off by the QM branch.\\ (b)~Stars with
a QM core and NM surface separated by a sharp interface (solid lines
to the left of dots) have masses up to about $1.6~\Msolar$.  For light
strange quarks $m_s\approx 150~\MeV$, color superconductivity
increases the maximum mass attained by the stars, but from a lower
starting point at $\De=0$.\\ (c)~Stars with a NM-QM mixed phase core
and NM surface (dashed lines) also have masses up to around
$1.6~\Msolar$ if the strange quark is heavy enough.\\ (d)~All
QM-containing stars with mass greater than about $1.6~\Msolar$ are
pure quark stars (solid lines to the right of dots).

Our overall conclusion is that in this range of values of the
bag constant, turning on color superconductivity allows 
hybrid stars to occur with masses up to around $1.6~\Msolar$. 
In Section~\ref{sec:conclusions} we will discuss in more detail 
whether this is compatible with recent observational data,
and how it should be used to interpret future observational data.

\begin{figure}[htb]
\begin{center}
\includegraphics[width=0.37\textwidth,angle=-90]{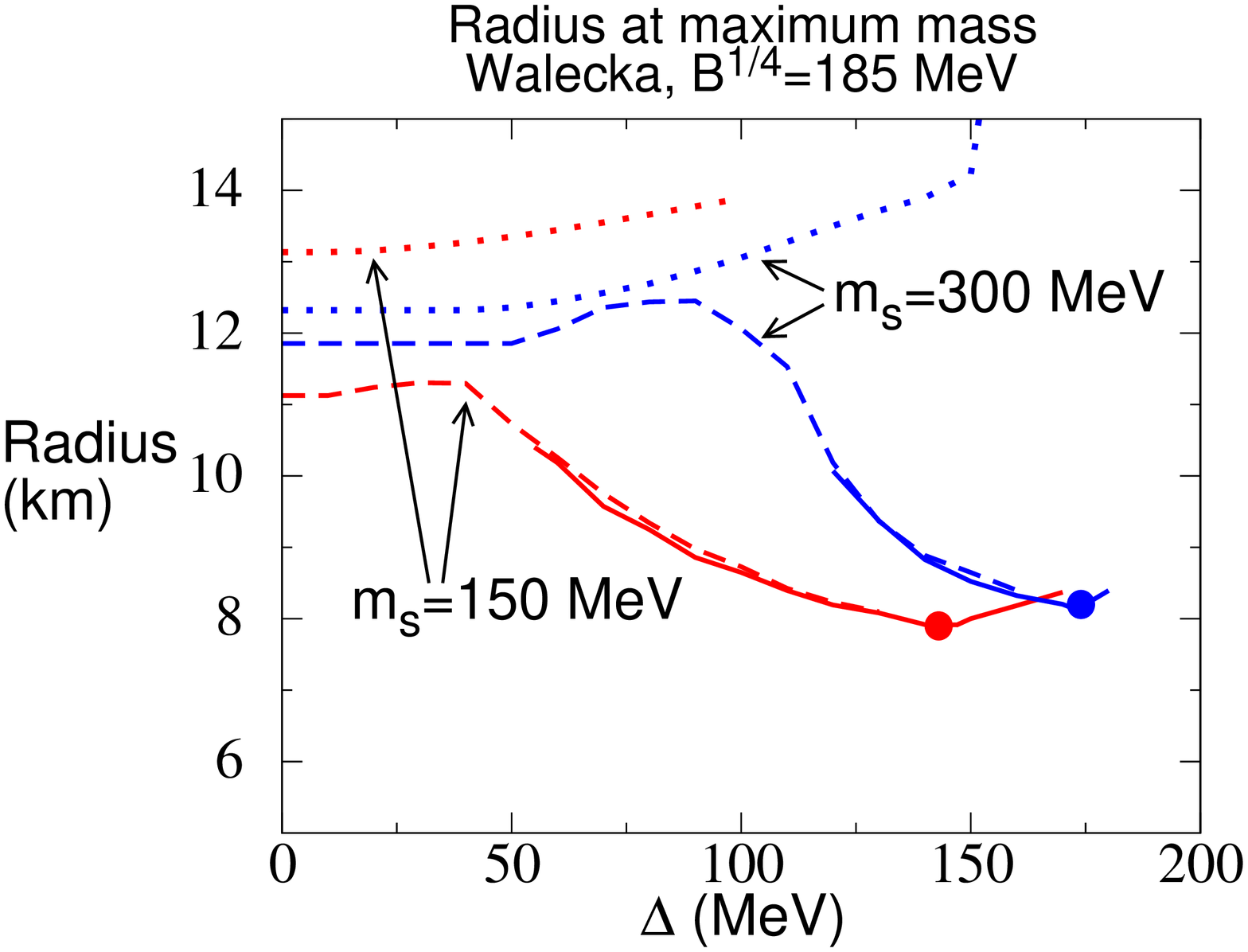}
\includegraphics[width=0.37\textwidth,angle=-90]{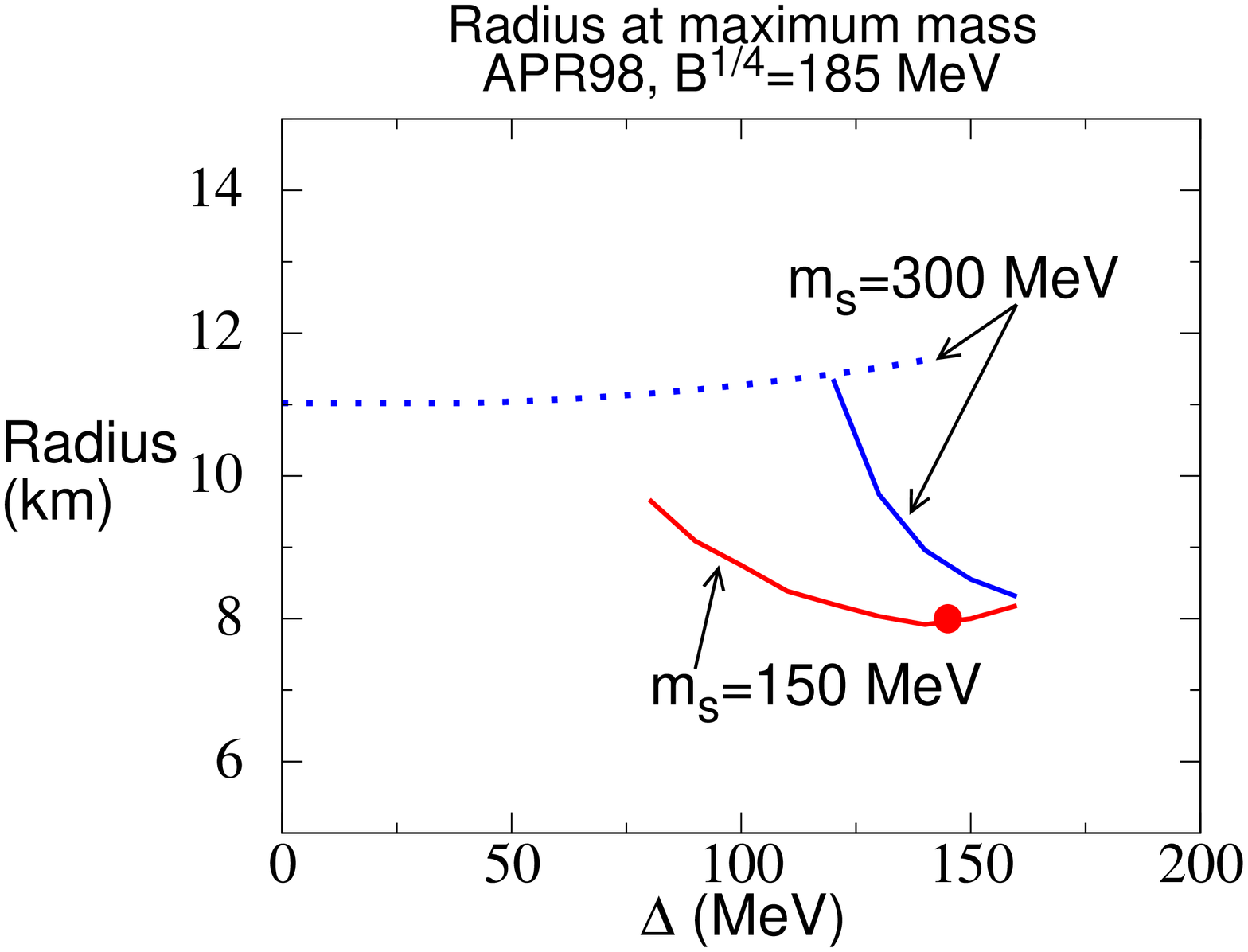}
\includegraphics[width=0.37\textwidth,angle=-90]{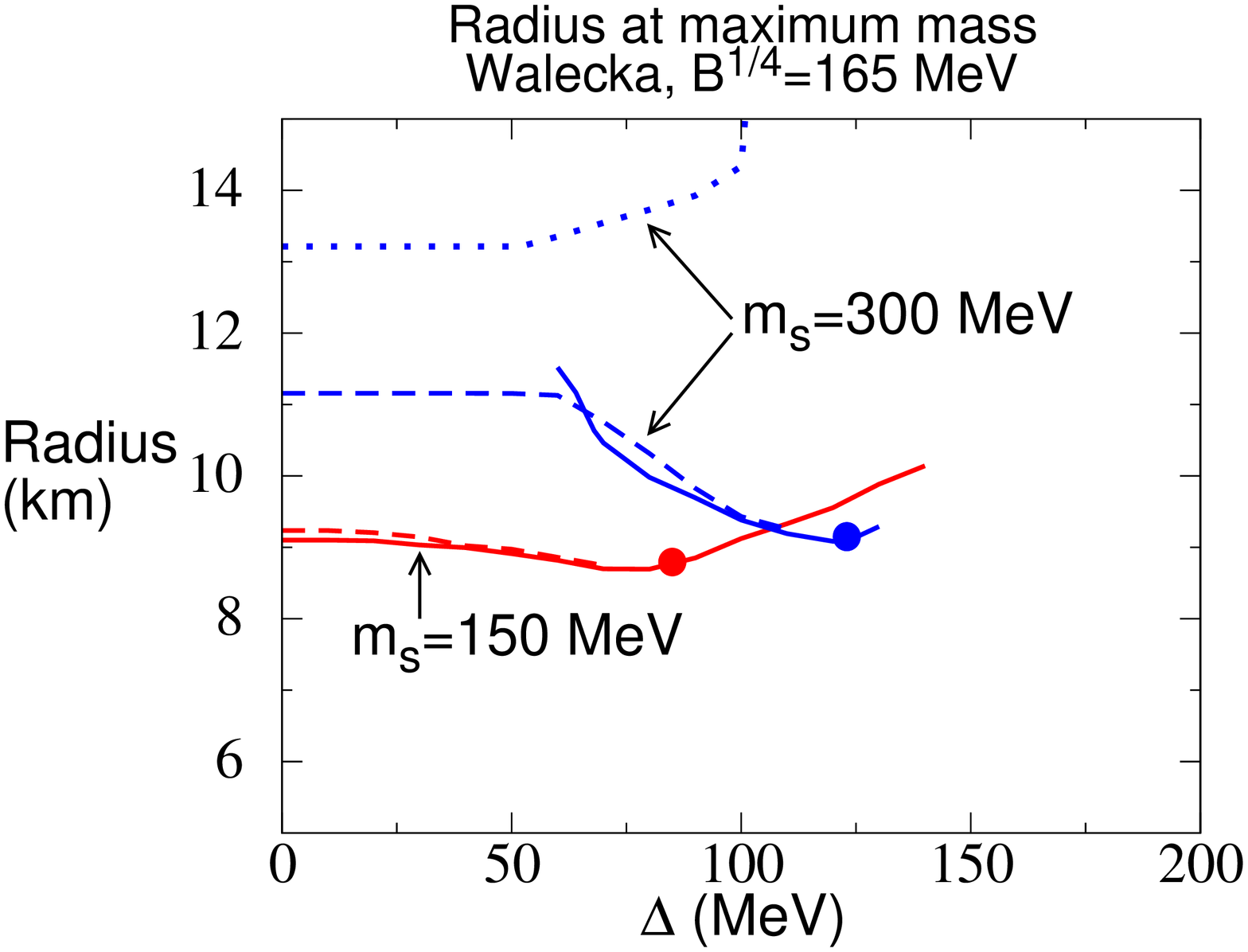}
\includegraphics[width=0.37\textwidth,angle=-90]{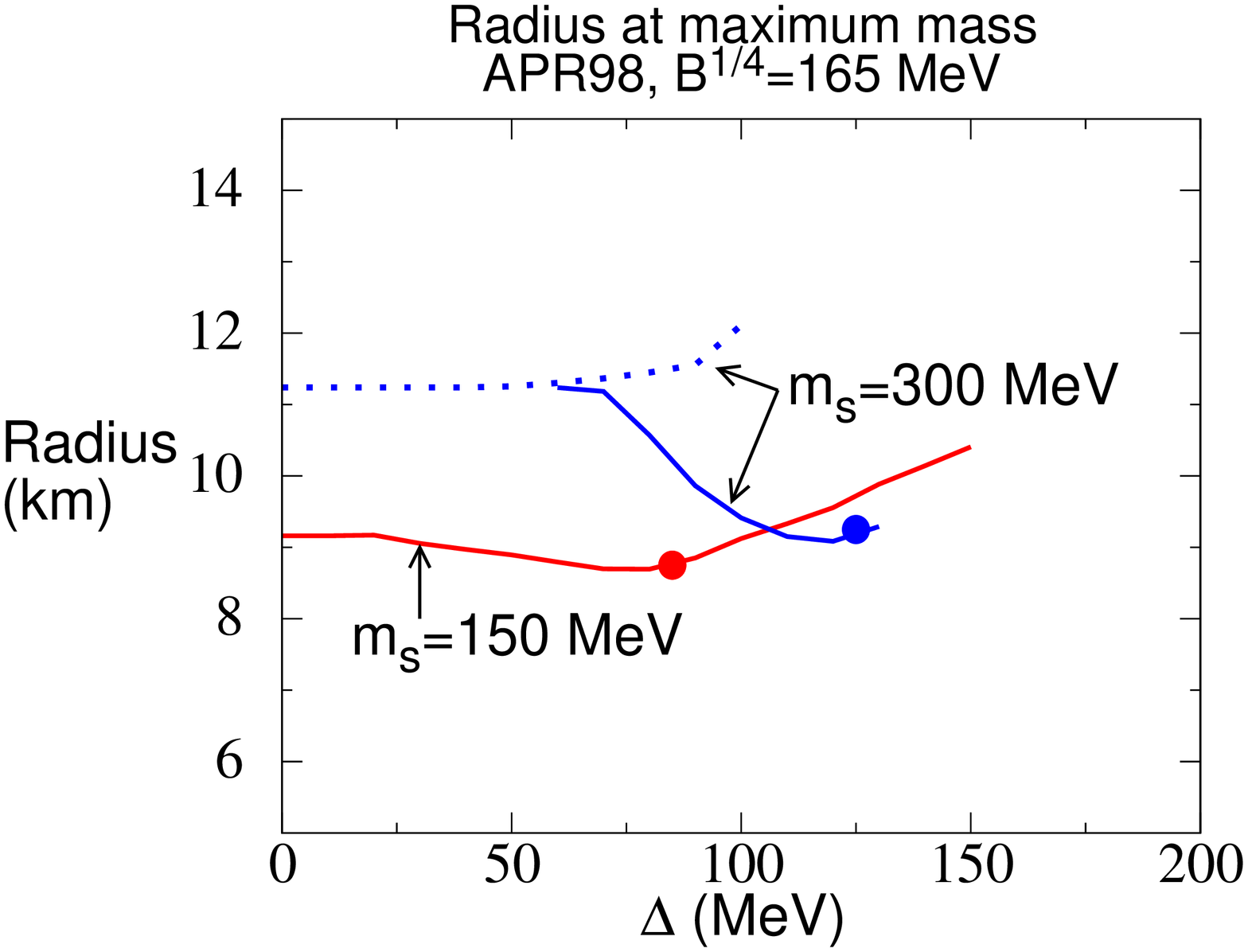}
\end{center}
\caption{ The radius of the star (in km) when it attains its maximum mass
as plotted in Fig.~\ref{fig:MaxM}. The upper plots are for
$B^{1/4}=185~\MeV$, the lower plots for $B^{1/4}=165~\MeV$. The left
plots are for nuclear matter described by the Walecka model (in which
case a mixed phase can be constructed).  The right plots are for the
APR98 nuclear equation of state.  Curves for strange quark mass $m_s=150$ and
$300~\MeV$ are shown.  The solid lines give the radius of stars
with QM only (to the right of the dot), 
or a QM core surrounded by NM (to the left of the dot). 
The dotted lines
show the heaviest pure NM star that occurs at the given gap $\De$.
The dashed lines are for stars that include a mixed phase. 
}
\label{fig:MaxM_R}
\end{figure}

\clearpage  

\subsection{Color superconductivity and compactness}
\label{sec:compactness}
By comparing the $\De=100~\MeV$ curves of Fig.~\ref{fig:massradius} 
with the UQM curve in Fig.~\ref{fig:simpleMR},
we see that superconductivity only has a moderate
effect on the radius of the star.
In Fig.~\ref{fig:MaxM_R} we investigate this issue more comprehensively,
by plotting the radii of the stars whose
masses appeared in Fig.~\ref{fig:MaxM}, ie the radius of the
heaviest star at each value of $\De$ for each equation of state.
This confirms that color superconductivity does not affect the radii
very strongly, but at low to medium gap $\De\lesssim 100~\MeV$
it tends to reduce them. This means that increasing $\De$ is
{\em not} simply equivalent to decreasing the bag constant,
in which case the mass and radius grow together as $1/\sqrt{B}$.
By comparing Fig.~\ref{fig:MaxM} with Fig.~\ref{fig:MaxM_R} we
see that when the maximum mass grows with $\De$, the radius
either decreases, or grows much more slowly.
Overall, the stars that contain QM have
radii between $8$ and $12~\km$. 

As one would expect from Fig.~\ref{fig:massradius}
the heaviest pure NM stars (dotted lines) have larger
radii $\gtrsim 11~\km$, since the QM stars replace the
NM stars at low radius (high central pressure).

\section{Conclusions}
\label{sec:conclusions}

We have seen that color superconductivity has a considerable effect on
the equation of state for quark matter. We can see from
\Eqn{OmegaQuark} and Fig.\ref{fig:massradius} (see also
Ref.~\cite{Lugones:2002ak}) that at values of the bag constant that
would normally preclude compact stars from containing any QM, a large
enough color-superconducting gap $\De$ can cancel out part of the bag
constant, allowing a stable hybrid star to occur.  In that sense,
turning up $\De$ has a similar effect to turning down $m_s$ or turning
down $B$, but there are differences (see below).

The important question for the interpretation of future observations
relates to the maximum mass of stars with quark matter in their core
but with a surface made of nuclear matter, as observed recently \cite{Cottam}
(see detailed discussion below).
Our overall conclusion is that, in the range of bag constants
and strange quark masses that we have studied, such stars
have a maximum mass around $1.6~\Msolar$.
This conclusion is based on our results
presented in Fig.~\ref{fig:MaxM}. The
top panels show that color superconductivity
allows QM-containing stars with a NM surface
to occur at $B^{1/4}=185~\MeV$ ($B=153~\MeV\!/\fm^3$), 
with masses up to about $1.5~\Msolar$.
The lower panels show that at $B^{1/4}=165~\MeV$ ($B=96~\MeV\!/\fm^3$)
color superconductivity allows QM+NM stars to exist at high $m_s$,
and at low $m_s$ it boosts their mass: at $m_s=150~\MeV$
a nuclear-surface star of mass $1.6~\Msolar$ is possible 
with $\De\sim 70~\MeV$, whereas the maximum mass would be 
$1.45~\Msolar$ without color superconductivity. 

It should be borne in mind that the QM+NM stars with masses
near $1.6~\Msolar$ are dominantly quark stars with a thin NM crust
(see Fig.~\ref{fig:profile}).
This is what one would expect, given that they occur just to the
left of the dots on the curves on Fig.~\ref{fig:MaxM} 
which mark the point at which the star becomes pure QM.
The NM$\to$QM transition in such stars occurs at 
very low pressure ($\lesssim 1~\MeV/\fm^3$) and at density
well below nuclear saturation density.

It is also interesting to note 
(see Fig.~\ref{fig:MaxM_R}) that at low $B$ and $m_s$
color superconductivity
increases the maximum mass without appreciably changing the
radius. This is different from the effect of changing the
bag constant, which increases the mass and radius together.

Finally, it is striking that
even under the circumstances where color superconductivity has
a noticeable effect on the maximum mass,
it only does so for gap $\De\gtrsim 50~\MeV$.
Smaller gaps have little effect.

We emphasize that we have had no difficulty in constructing stars that
are compatible with the recent results of Cottam, Paerels, and Mendez
\cite{Cottam}, who obtained a observational $M$-$R$ curve by measuring
the redshift of emission lines from highly ionized Oxygen and Iron in
X-ray bursts from EXO0748-676. At radii of $8$-$12~\km$, their results
suggest that the compact star mass is $1.2$-$1.8\,\Msolar$.  Most of
our stars fall in this range. Another constraint on the mass-radius of
compact objects arise from recent observations of thermal radiation
from RXJ185635-3754, an isolated neutron star. This combined with an
accurate measurement of the distance to this object provide some
information about its radius. However, since the spectrum is not quite
black-body, the inferred radius depends on theoretical models employed
to describe the objects atmosphere
\cite{Pons:2001px,Walter:2002uq}. For this reason, the constraint from
RXJ185635-3754 is weak and not as compelling as those derived from
EXO0748-676. A simple black-body fit yields radii that are small
$R\sim 5$ km \cite{Drake:2002bj}. In our study here we were unable to
construct stars with such small radii.  Atmosphere models which best
fit the observed data favor a large radius.  In a recent article,
Walter and Lattimer find that these model studies indicate that
$R=11.4 \pm 2$ km and $M=1.7 \pm 0.4 M_{\odot}$
\cite{Walter:2002uq}. The superconducting quark stars constructed in
this work are marginally compatible with these results. These larger
values for the mass and radius require a small bag constant, a large
$\De$ and small $m_s$.

We have shown that color superconductivity allows hybrid stars to have
masses up to about $1.6~\Msolar$: does this mean that a definitive 
observation of 
a significantly heavier compact star would rule out quark matter?
Our calculation could be improved in many ways (discussed below),
so the upper limit we quote has theoretical errors that can only
be roughly estimated. From Fig.~\ref{fig:MaxM} we see that varying $m_s$ 
over the plausible range changes
the maximum mass by less than $0.1~\Msolar$, and the
mass rises from about $1.5~\Msolar$
at $B^{1/4}=185~\MeV$ to about $1.6~\Msolar$ at $B^{1/4}=165~\MeV$.
Given this level of theoretical uncertainty, it
seems that a definitive observation of a star with
$M\gtrsim 1.8~\Msolar$ would be difficult to explain in terms of 
hybrid QM+NM stars
without invoking an even lower bag constant, with the danger that
nuclear matter will be rendered unstable against two-flavor quark
matter.

There are many ways in which this line of inquiry could be
pursued further.
(1)~
We treated the quarks as free, with a 
color superconducting gap at their Fermi surface.
We did not include perturbative corrections to the equation of
state \cite{pertEoS}, in other words, we set 
the strong coupling constant $\al_s=0$.
It would be useful to perform
such calculations for CFL quark matter, and see how robust our conclusions
are against variation in $\al_s$.
(2)~
We allowed two phases of quark matter, unpaired and CFL, and we did
not include the two-flavor color superconducting ``2SC'' phase.
Although this phase is generally unfavored \cite{AR-02,Steiner:2002gx},
it is just possible that there is a narrow range of $m_s$ 
in which it can occur. Another competitor is the
crystalline phase \cite{firstLOFF,Bowers:2002xr}.
It would be interesting to include these additional phases
in our calculations.
(3)~
We have used a general bag-model expression for
the free energy of the quark matter, neglecting any possible
density-dependence of the strange quark mass
and color superconducting gap. We have also assumed that the bag constant
takes the same value in all the quark matter phases, neglecting any
differences of ground state energy between them.
It would be interesting to use an NJL model to calculate the
quark matter equation of state, and thereby include these phenomena.
One could also use a bag constant with a
phenomenological density dependence \cite{Burgio:2001mk}.

It is very encouraging that observations of the radii of compact stars
are becoming more precise.  There have been significant developments
in the theory of dense quark matter in the last few years, and we look
forward to seeing whether the observed properties of compact stars are
compatible with (or even require) the presence of the exotic phases of
quark matter that are being so widely discussed.

\vspace{3ex}
{\samepage 
\begin{center} {\bf Acknowledgments} \end{center}
\nopagebreak
We have had useful discussions with K. Rajagopal and V. Pandharipande.
We thank Dick Silbar for a careful reading of the manuscript. The
research of MGA is supported in part by the UK PPARC.  The research of
SR is supported in part by funds provided by the U.S. Department of
Energy (D.O.E.)  under cooperative research agreement
DF-FC02-94ER40818 and the D.O.E. contract W-7405-ENG-36.  }


\begin{thebibliography}{99}


\bibitem{Barrois}
B.~C.~Barrois,
Nucl.\ Phys.\ B {\bf 129}, 390 (1977).
S. Frautschi, Proceedings of workshop on hadronic matter at extreme density,
Erice 1978.
B. Barrois, ``Nonperturbative effects in dense quark matter'',
Cal Tech PhD thesis, UMI 79-04847-mc (1979).

\bibitem{BailinLove}
D.~Bailin and A.~Love,
Phys.\ Rept.\  {\bf 107}, 325 (1984).

\bibitem{ARW1}
M.~Alford, K.~Rajagopal and F.~Wilczek,
Phys.\ Lett.\ B {\bf 422}, 247 (1998)
[hep-ph/9711395].

\bibitem{RappETC}
R. Rapp, T. Sch\"afer, E. V. Shuryak and M. Velkovsky, 
Phys.\ Rev.\ Lett.\  {\bf 81}, 53 (1998)
[hep-ph/9711396].

\bibitem{CFL}
M. Alford, K. Rajagopal and F. Wilczek, 
Nucl.\ Phys.\ B {\bf 537}, 443 (1999)
[hep-ph/9804403].

\bibitem{Reviews}
M.~G.~Alford,
Ann.\ Rev.\ Nucl.\ Part.\ Sci.\  {\bf 51} (2001) 131
[hep-ph/0102047].
%
K.~Rajagopal and F.~Wilczek,
hep-ph/0011333.
%
T.~Sch\"afer and E.~V.~Shuryak,
Lect.\ Notes Phys.\  {\bf 578} (2001) 203
[nucl-th/0010049].
%
D.~K.~Hong,
Acta Phys.\ Polon.\ B {\bf 32} (2001) 1253
[hep-ph/0101025].
%
S.~D.~Hsu,
hep-ph/0003140.
%
D.~H.~Rischke and R.~D.~Pisarski,
nucl-th/0004016.


\bibitem{ichep}
M.~G.~Alford,
hep-ph/0209287.

\bibitem{Burgio:2002sn}
G.~F.~Burgio, M.~Baldo, P.~K.~Sahu and H.~J.~Schulze,
Phys.\ Rev.\ C {\bf 66}, 025802 (2002)
[nucl-th/0206009].

\bibitem{Schertler:2000xq}
K.~Schertler, C.~Greiner, J.~Schaffner-Bielich and M.~H.~Thoma,
Nucl.\ Phys.\ A {\bf 677}, 463 (2000)
[astro-ph/0001467].

\bibitem{Steiner:2000bi}
A.~Steiner, M.~Prakash and J.~M.~Lattimer,
Phys.\ Lett.\ B {\bf 486}, 239 (2000)
[nucl-th/0003066].



\bibitem{Lugones:2002ak}
G.~Lugones and J.~E.~Horvath,
hep-ph/0211070.

\bibitem{APR98}
A. Akmal, V.R. Pandharipande, D.G. Ravenhall,
Phys.Rev. C58 1804 (1998)
[nucl-th/9804027].

\bibitem{NV}
J. Negele and D. Vautherin, Nucl. Phys. {\bf A207}, 298 (1973).

\bibitem{BPS} G. Baym, C. Pethick, D. Sutherland,
Astrophys. J. {\bf 170}, 299 (1971).

\bibitem{GBOOK}
N. K. Glendenning, {\it Compact Stars, Nuclear Physics, Particle Physics \\
and General Relativity }, (Springer-Verlag, New York, 1997)

\bibitem{SW}
B. Serot and J. D. Walecka, {\it Advances in Nucl. Physics}, {\bf 16}, 
edited by J. W. Negele and E. Vogt (Plenum, New York, 1986)

\bibitem{neutrality}
K.~Rajagopal and F.~Wilczek,
Phys.\ Rev.\ Lett.\  {\bf 86}, 3492 (2001)
[hep-ph/0012039].


\bibitem{effectivetheory}
D. Son, M. Stephanov, 
Phys.\ Rev.\ D {\bf 61}, 074012 (2000)
[hep-ph/9910491],
erratum 
Phys.\ Rev.\ D {\bf 62}, 059902 (2000)
[hep-ph/0004095].

\bibitem{BedaqueSchaefer}
P.~F.~Bedaque and T.~Sch\"afer,
Nucl.\ Phys.\ A {\bf 697} (2002) 802
[hep-ph/0105150].

\bibitem{KaplanReddy}
D.~B.~Kaplan and S.~Reddy,
Phys.\ Rev.\ D {\bf 65}, 054042 (2002)
[arXiv:hep-ph/0107265].


\bibitem{Glendenning:1992vb}
N.~K.~Glendenning,
Phys.\ Rev.\ D {\bf 46}, 1274 (1992).

\bibitem{ARRW}
M.~G.~Alford, K.~Rajagopal, S.~Reddy and F.~Wilczek,
hep-ph/0105009.



\bibitem{TOV}
R. Tolman, Phys. Rev. {\bf 55}, 364 (1939);
J. Oppenheimer and G. Volkoff, Phys. Rev. {\bf 55}, 374 (1939).

\bibitem{Cottam}
J.~Cottam, F.~Paerels, M.~Mendez, Nature {\bf 420}, 51 (2002).

\bibitem{Lattimer:2000nx}
J.~M.~Lattimer and M.~Prakash,
Astrophys.\ J.\  {\bf 550}, 426 (2001)
[astro-ph/0002232].

\bibitem{Pons:2001px} J.~A.~Pons, F.~M.~Walter, J.~M.~Lattimer,
M.~Prakash, R.~Neuhauser and P.~h.~An,
Astrophys.\ J.\  {\bf 564}, 981 (2002)
[astro-ph/0107404].

\bibitem{Walter:2002uq}
F.~M.~Walter and J.~Lattimer,
Astrophys.\ J. \ {\bf 576}, L145-L148 (2002)
[astro-ph/0204199].

\bibitem{Drake:2002bj}
J.~J.~Drake {\it et al.},
Astrophys.\ J.\  {\bf 572}, 996 (2002)
[astro-ph/0204159].


\bibitem{pertEoS}
E.~Farhi and R.~L.~Jaffe,
Phys.\ Rev.\ D {\bf 30}, 2379 (1984).
%
B.~A.~Freedman and L.~D.~McLerran,
Phys.\ Rev.\ D {\bf 16}, 1130 (1977);
%
Phys.\ Rev.\ D {\bf 16}, 1147 (1977);
%
Phys.\ Rev.\ D {\bf 16}, 1169 (1977);
%
Phys.\ Rev.\ D {\bf 17}, 1109 (1978).
%
V.~Baluni,
Phys.\ Rev.\ D {\bf 17}, 2092 (1978).

\bibitem{AR-02}
M.~Alford and K.~Rajagopal,
JHEP {\bf 0206} (2002) 031
[hep-ph/0204001].


\bibitem{Steiner:2002gx}
A.~W.~Steiner, S.~Reddy and M.~Prakash,
hep-ph/0205201.

\bibitem{firstLOFF}
M.~Alford, J.~Bowers and K.~Rajagopal,
Phys.\ Rev.\ D {\bf 63}, 074016 (2001)
[hep-ph/0008208].

\bibitem{Bowers:2002xr}
J.~A.~Bowers and K.~Rajagopal,
hep-ph/0204079.

\bibitem{Fraga}
E.~S.~Fraga, R.~D.~Pisarski and J.~Schaffner-Bielich,
Phys.\ Rev. {\bf D63}, 121702 (2001)
[hep-ph/0101143].


\bibitem{Burgio:2001mk}
G.~F.~Burgio, M.~Baldo, P.~K.~Sahu, A.~B.~Santra and H.~J.~Schulze,
Phys.\ Lett.\ B {\bf 526}, 19 (2002)
[astro-ph/0111440].

\end{thebibliography}
\end{document}